\documentclass[prb,aps,preprint]{revtex4}
\usepackage{graphicx}

\begin{document}

\title{Will Zigzag Graphene Nanoribbon Turn to Half Metal under Electric Field?}

\author{Er-Jun Kan}

\author{Zhenyu Li}

\author{Jinlong Yang }

\thanks{Corresponding author. E-mail: jlyang@ustc.edu.cn}

\author{J. G. Hou }

\affiliation{Hefei National Laboratory for Physical Sciences at
     Microscale,  University of Science and Technology of
     China, Hefei,  Anhui 230026, People's Republic of China}

\date{\today}

\begin{abstract}
At B3LYP level of theory, we predict that the half-metallicity in
zigzag edge graphene nanoribbon (ZGNR) can be realized when an
external electric field is applied across the ribbon. The critical
electric field to induce the half-metallicity decreases with the
increase of the ribbon width. Both the spin polarization and
half-metallicity are removed when the edge state electrons fully
transferred from one side to the other under very strong electric
field. The electric field range under which ZGNR remain
half-metallic increases with the ribbon width. Our study
demonstrates a rich field-induced spin polarization behavior,
which may leads to some important applications in spinstronics.
\end{abstract}

\maketitle Since its isolation by mechanical
exfoliation\cite{ref1}, graphene, a single graphite layer, has
attracted a broad research interest. Because of its high carrier
mobility and huge coherence distance at room
temperature\cite{ref2,ref3} it has been expected to be a very
promising candidate of the future electronics materials. Along
this direction, graphene nanoribbon (GNR), which is a graphite
layer terminated in one direction with a specific width, were
synthesized\cite{ref2,ref3,ref4}. It is now well known that zigzag
edge GNRs (ZGNRs) are semiconductors with two localized electronic
edge states\cite{ref5,ref6,ref7,ref8}. These two states are
ferromagnetically ordered, and they antiferromagnetically coupled
each other.

The spin degree of freedom of ZGNRs is very important considering
their possible application in spintronics\cite{ref9,ref10}, where
it is essential to realize electron transport through only one
spin channel. With one metallic spin component and another
semiconducting or insulating spin channel, half metal is an ideal
material for spintronics. Based on density functional theory (DFT)
calculations with the local density approximation (LDA), Son et
al\cite{ref11}. predicted that ZGNRs become half-metallic when an
external transverse electric field is applied, which opens the
possibility of the spintronics application for graphene.

Recently, Rudberg et al\cite{ref12}. revisited the problem of the
ZGNR electronic structure under electric field. They argued that
although ZGNRs are spin-selective semiconductor, they never show
half-metallicity if nonlocal exchange interaction is considered by
using B3LYP functional. However, their study is limited to a
specific ZGNR with 8 zigzag chains (8-ZGNR) and of finite length.
Therefore, it is still an open question if ZGNR will turn to half
metal under electric field.

\begin{figure}
 \caption{(Color online) The structures of zigzag graphene nanoribbons, green balls are C atoms,
  blue balls are H atoms. The rectangle drawn with solid lines denotes the unit cell,
  and the arrow line represents the direction of external electric fields.}
  \label{fig1}   \includegraphics{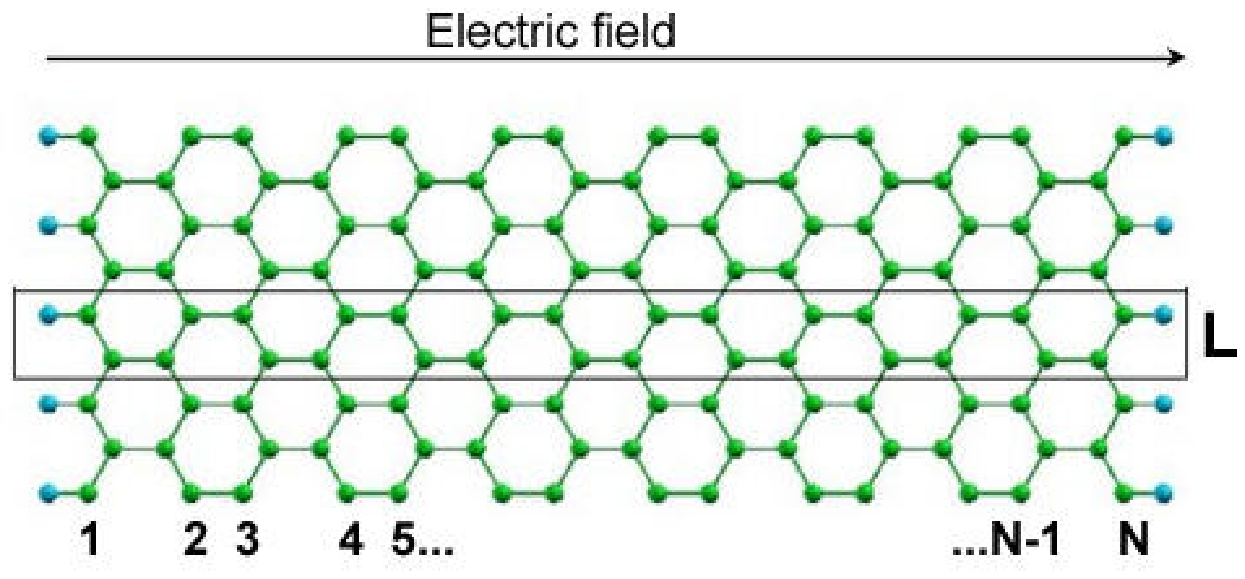}
  \end{figure}

In this letter,to clarify this issue, we report a careful study on
the electronic structures of n-ZGNRs (n=5, 6, 7, 8, 9, 10, 12, and
14) with B3LYP functional. In our model, n-ZGNR is flat in x-y
plan, with n zigzag chains along y direction. The edges of ZGNRs
are saturated by hydrogen atoms. Periodic boundary condition (PBC)
is used to consider ZGNRs with infinite length, just as shown in
Fig.~\ref{fig1}. Our calculations were carried out with the
CRYSTAL03 package\cite{ref13}, which utilizes atom-centered
Gaussian basis sets\cite{ref14}. All-electron basis sets were
adopted for both C\cite{ref15} and H.\cite{ref16} A reciprocal
space sampling was made on a Monkhorst-Pack grid with a shrinking
factor sufficient to converge the total energy to within eV per
unit cell. A modified Broyden scheme\cite{ref17}, following the
method proposed by Johnson\cite{ref18}, was applied in the SCF
iteration.

\begin{figure}
 \caption{(Color online) (a) Spin-up (blue) and spin-down (red) 8-ZGNR band gaps against external electric fields.
 (b) Spin-up (blue) and spin-down (red) 8-ZGNR band structure with E = 0.65 V/{\AA}.
 (c) The spin densities of 8-ZGNR under different external electric fields, red for positive values and blue for negative. }
  \label{fig2}
  \includegraphics[scale=0.7]{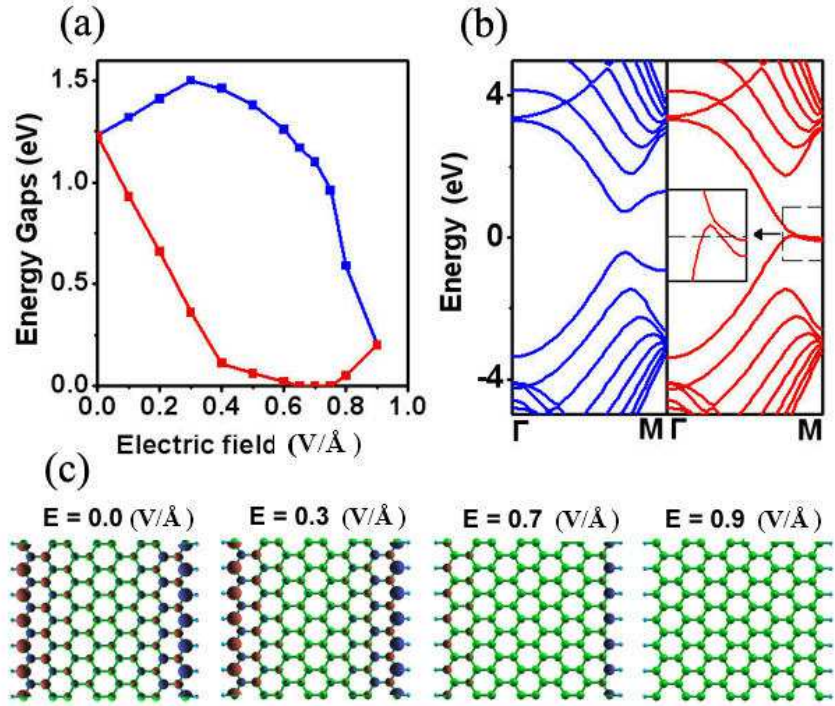}
  \end{figure}

We first consider the 8-ZGNR studied by Rudberg et al\cite{ref12}.
As shown in Fig.~\ref{fig2} (a), without electric field, the
energy gaps for both spin channels are 1.23 eV, very close to the
1.34 eV for ribbon of length 7.1 nm\cite{ref12}. With the increase
of the external electric field, the spin-down band gap decreases
rapidly, and becomes zero when the electric field strength reaches
0.65 V/{\AA}. On the other hand, the spin-up channel remains
semiconducting under all external electric fields. Therefore,
B3LYP also predict the electric field induced half-metallicity for
the infinite long 8-ZGNR. However, the critical electric field
obtained here is much larger than previously predicted by LDA
(about 0.2 V/{\AA})\cite{ref11}. Another important new feature
revealed by our calculation is that the half-metallicity will be
destroyed by a too strong electric field. As shown in Fig. 2 (a),
the metallic character in the spin-down channel disappears when
the electric field reaches 0.8 V/{\AA}, Therefore, the answer to
the title question is yes, and 8-ZGNR will turn to half metal
under a limited range of external electric fields.

To understand the electric field response of 8-ZGNR, we plot its
spin densities ($\rho_{\alpha}$(r)- $\rho_{\beta}$(r)) of
8-ZGNR in Fig. 2 (c). Without electric field, we obtained a spin
density very similar to the previous results\cite{ref11,ref12}. It
distributes on all carbon atoms, and decays from the edges to the
middle. Each spin-up and spin-down density is mainly distributed
at one side of the ribbon. By applying a 0.3 V/{\AA} external
electric field, the spin density in the middle of the ribbon is
reduced, but the two edges are only little affected.  When the
electric field increases further and the ribbon becomes half
metal, the spin density at the two edges is greatly reduced, and
there is almost no spin density on the middle part. Finally, when
the field reaches 0.9 V/{\AA}, the system becomes
spin-unpolarized. Rudberg et al.\cite{ref12} found the
magnetization decrease with the increase of electric field too.
However, in their limited length case, the spin density disappears
first from the two ends of the ribbon, and some magic patterns
formed along the y direction under some fields, which directly
leads to the half-metallicity unavailable for ribbons of finite
length.

\begin{figure}
 \caption{(Color online) Project density of states (PDOS) of 8-ZGNR under external electric field of (a) 0.0 V/{\AA},
 (b) 0.9 V/{\AA}, positive for spin-up, negative for spin-down. (c) N-ZGNR band gaps against external electric fields for
  n= 7, 9 , 10, 12, 14,  the line with squares represents spin-up channel, and filled circles for spin-down one.
  (d) The critical electric fields (E$_t$) to achieve half-metallicity, and the range of electric fields strength
  ( from E$_t$ to E$_t$ +$\Delta$E) to keep half-metallicity for ribbons with n= 5, 6, 7, 8, 9, 10, 12, 14. }
  \label{fig3}
  \includegraphics[scale=0.5]{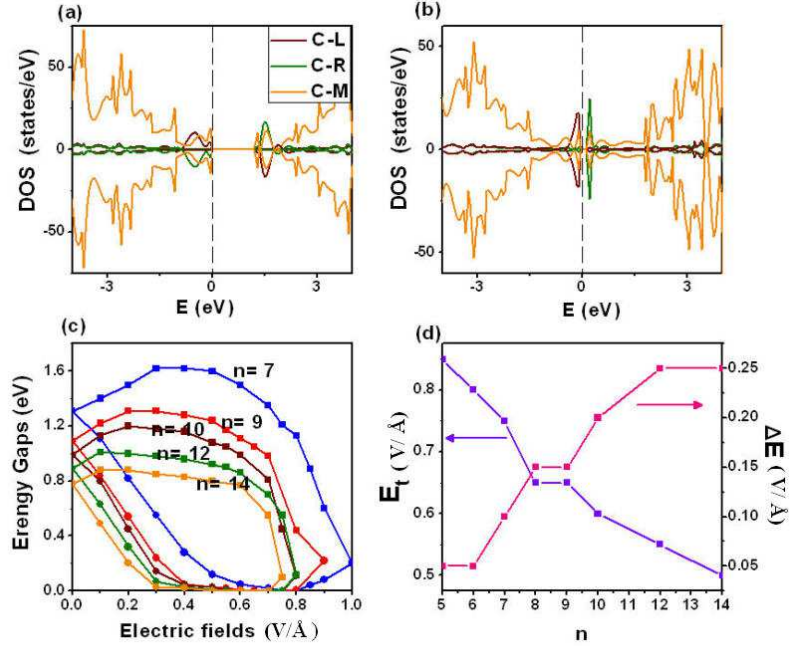}
  \end{figure}

As shown in Fig.~\ref{fig3} (a) and (b), the projected density of
states (PDOS) of 8-ZGNR is obtained by projecting electronic bands
to three parts: atomic orbitals at the left/right C atom chain
(C-L/C-R), and other C atoms (C-M). As shown in Fig.~\ref{fig3}
(a), without external electric field, the two edge states are
degenerate and with opposite spin directions. The degeneration of
the two edge states is removed when external electric field is
applied, due to the different electrostatic potentials at the two
sides. The energy difference between these two edge states changes
with the electric field strengths. When the occupied manifest of
the left edge state meets the unoccupied manifest of the right
edge state, changer transfer occurs and the spin-down channel
becomes metallic. Under an electric field of 0.9 V/{\AA}, all
electrons of the right edge state have been transferred to the
left edge, making the spin density quenched. Thus, the
half-metallicity is closely related to the electron transfer from
right to the left.

It is interesting to check whether electric field can reduce
half-metallicity for other ribbons with different widths. As shown
in Fig. 3 (c), our calculations predict that all n-ZGNRs (n=5, 6,
7, 9, 10, 12, and 14) can display half-metallic behavior under
suitable external field. The critical electric field to achieve
half-metallicity decreases with the increase of ribbon width. This
can be easily understood by the different electric field
requirements to generate the same electrostatic potential
difference between the left and the right side of
n-ZGNR\cite{ref11}. As shown in Fig.~\ref{fig3} (d), n-ZGNR
remains half-metallic only at a limited range of electric field.
With the increase of the ribbon widths, the corresponding field
range increases too. The reason is that it is easier for wider
ZGNRs to induce bands crossing in the spin-down channel by
electric field. Band crossing leads to charge transfer. Once the
electron transfer removes the spin polarization, the
half-mtallicity has already been destroyed.

In summary, we have performed hybrid functional calculations with
PBC to study the electronic structures of ZGNRs under external
electric field. We find that the nonlocal exchange interaction
does not remove half-metallicity of ZGNRs, and the contrary
prediction by Rudberg et al\cite{ref12}. should be an effect of
finite length. The half-metallic behaviors are correlated with the
partial electron transfer between the two edges of the ribbons and
the band crossing caused by the additive electric potential
difference under the electric field. A full edge state electron
transfer removes spin polarization, and thus also the
half-metallicity. The electric field range at which ZGNR remains
half-metallic increases with the ribbon width. This new prediction
may lead to some important applications of ZGNR in spintronics,
such as switches and sensors.

This work is partially supported by the National Natural Science
Foundation of China (50121202, 20533030, 10474087), by National
Key Basic Research Program under Grant No. 2006CB922004, by the
USTC-HP HPC project, and by the SCCAS and Shanghai Supercomputer
Center.

\newpage

\end{document}